\documentclass[a4paper,11pt]{article}
\usepackage{pos}
\usepackage{wrapfig}
\usepackage{sidecap}
\usepackage{tikz}
\usetikzlibrary{arrows.meta, automata, calc, positioning, quotes}

\makeatletter
\g@addto@macro\bfseries{\boldmath}
\makeatother

\title{Continuum limit of baryon-baryon scattering \\ with SU(3) flavor symmetry}
\ShortTitle{Continuum limit of baryon-baryon scattering with SU(3) flavor symmetry}

\author*[a,1]{Jeremy R.\ Green}
\author[b,c,d]{Andrew D.\ Hanlon}
\author[e]{Parikshit M.\ Junnarkar}
\author[c,d,f]{Hartmut~Wittig}

\affiliation[a]{Theoretical Physics Department, CERN, 1211 Geneva 23, Switzerland}
\note{Present address: School of Mathematics and Hamilton Mathematics Institute, Trinity College, Dublin 2, Ireland}

\affiliation[b]{Physics Department, Brookhaven National Laboratory, Upton, New York 11973, USA}
\affiliation[c]{Helmholtz-Institut Mainz, Johannes Gutenberg-Universität, 55099 Mainz, Germany}
\affiliation[d]{GSI Helmholtzzentrum für Schwerionenforschung, 64291 Darmstadt, Germany}
\affiliation[e]{Institut für Kernphysik, Technische Universität
Darmstadt, Schlossgartenstraße 2, 64289 Darmstadt, Germany}
\affiliation[f]{PRISMA Cluster of Excellence and Institut für Kernphysik, University of Mainz, Becher Weg 45, D-55099 Mainz, Germany}

\emailAdd{green@maths.tcd.ie}
\emailAdd{ahanlon@bnl.gov}
\emailAdd{parikshit@theorie.ikp.physik.tu-darmstadt.de}
\emailAdd{hartmut.wittig@uni-mainz.de}

\abstract{We report a study of the scattering of two octet baryons
  using lattice QCD. The baryon-baryon spectrum is computed using
  distillation on eight lattice ensembles spanning six lattice
  spacings and multiple volumes, all at the SU(3) flavor symmetric
  point with $m_\pi=m_K\approx 420$~MeV. Using finite-volume
  quantization conditions, we determine the scattering phase shift and
  the presence of bound states. Focusing on the $H$ dibaryon, our
  results show large discretization effects: in the continuum, the
  binding energy is $B_H=4.56\pm1.13\pm0.63$~MeV, whereas on our
  coarsest lattice spacing this is larger by a factor of about 7.5. We
  also present preliminary results for a $D$-wave phase shift and for
  the spectrum in the nucleon-nucleon ${}^1S_0$ channel.}

\FullConference{%
 The 38th International Symposium on Lattice Field Theory, LATTICE2021
  26th-30th July, 2021
  Zoom/Gather@Massachusetts Institute of Technology
}

\newcommand{\cO}{\mathcal{O}}

\begin{document}
\maketitle

\section{Introduction}

It is very computationally challenging to study multibaryon systems
using lattice QCD. Because of this, past calculations had to make the
assumption that discretization effects are small. In these
proceedings, we present calculations done at a single SU(3)-symmetric
quark mass point, covering a wide range of lattice spacings and
several volumes. This allows us to perform the first systematic study
of discretization effects in a multibaryon system.

In Section~\ref{sec:setup}, we briefly describe our calculation and in
Section~\ref{sec:Hdib}, we summarize a study of the $H$ dibaryon that was
already reported in Ref.~\cite{Green:2021qol}. The talk at Lattice
2021 was based on the first version of \cite{Green:2021qol}. Since
then, we have revised the analysis by including an additional ensemble
and choosing fit regions for determining the spectrum in a more
conservative way. These proceedings are based on the revised
analysis. For a more complete discussion of the setup of the
calculation and the $H$ dibaryon study, we refer the reader to
Ref.~\cite{Green:2021qol}.

The following two sections describe preliminary analyses that extend
our work to additional systems: the ${}^1D_2$ partial wave
(Section~\ref{sec:Dwave}) and the SU(3) 27-plet
(Section~\ref{sec:27}). Finally, our conclusions are in
Section~\ref{sec:conclu}.

\section{Lattice setup}
\label{sec:setup}

\begin{wrapfigure}{R}{0.5\textwidth}
  \centering
  \includegraphics[width=0.5\textwidth]{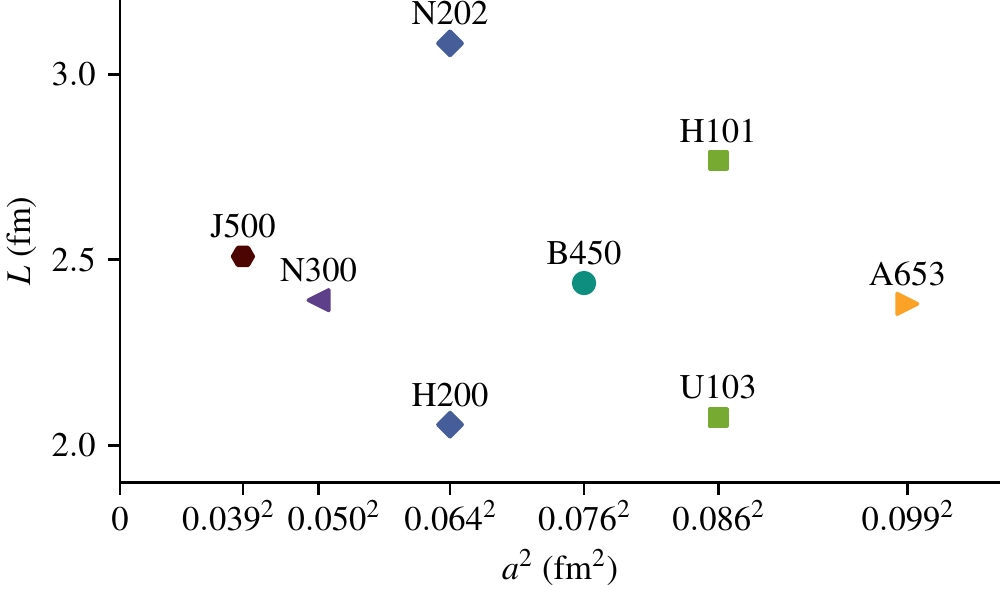}
  \caption{Box size $L$ and lattice spacing $a$ for the ensembles used
    in this work.}
  \label{fig:ensembles}
\end{wrapfigure}

This calculation is based on eight ensembles generated by
CLS~\cite{Bruno:2014jqa}, with three degenerate quarks that have a
mass set to the average of the physical $u$, $d$, and $s$ quark
masses, corresponding to $m_\pi=m_K\approx 420$~MeV. The ensembles
span six lattice spacings and a range of volumes, as shown in
Fig.~\ref{fig:ensembles}.

Our analysis is based on finite-volume spectroscopy and quantization
conditions. In a given symmetry sector, we compute a matrix of
two-point correlation functions,
\begin{equation}
  C_{ij}(t) \equiv \langle \cO_i(t) \cO_j^\dagger(0) \rangle,
\end{equation}
where the two-baryon interpolating operators $\{\cO_j\}$ are formed
from linear combinations of products of momentum-projected
single-baryon interpolators. In addition to definite flavor, total
momentum $\vec P$, and irreducible representation of the little group
of $\vec P$, the operators also have a definite two-baryon spin.

To determine the spectrum, we solve a generalized eigenvalue problem,
$C(t_D)v_n=\lambda_nC(t_0)v_n$, and use the eigenvectors to construct
an approximately diagonalized correlator matrix $\tilde C(t)$ with
diagonal entries approximately proportional to $e^{-E_n t}$ at large
$t$. We form the ratio of $\tilde C_n(t)$ to a product of two
single-baryon correlators and perform fits to obtain the shift
$\Delta E_n$ from the corresponding noninteracting level.

Finite-volume quantization conditions~\cite{Luscher:1990ux,
  Rummukainen:1995vs, Briceno:2013lba, Briceno:2014oea} provide a
relation between the spectrum and the baryon-baryon scattering
amplitude. We use a form similar to Ref.~\cite{Morningstar:2017spu},
which says that the spectrum is given by solutions of
\begin{equation}
  \det\left[ \tilde K^{-1}(p^2) - B(p^2) \right] = 0,
\end{equation}
where $\tilde K$ contains the scattering amplitude and $B$ depends on
the volume, $\vec P$, and irrep. In this work, our preferred kinematic
variable is the center-of-mass momentum
$p^2\equiv (E_\text{cm}/2)^2-m_B^2$.

We study the scattering of two octet baryons. The flavor content of
this baryon-baryon system belongs to one of five SU(3) irreps:
\begin{equation}
  \boldsymbol{8}\otimes\boldsymbol{8} =
  (\boldsymbol{1}\oplus\boldsymbol{8}\oplus\boldsymbol{27})_S
  \oplus (\boldsymbol{8}\oplus\boldsymbol{10}\oplus\boldsymbol{\overline{10}})_A,
\end{equation}
where the subscripts denote irreps appearing in the symmetric and
antisymmetric products. The simplest irreps to study are the singlet
and septenvigintuplet, since they appear only in the symmetric
product, meaning that even partial waves have spin zero and do not
couple to other partial waves. In particular, our initial focus is on
the $H$ dibaryon, which appears in the singlet ${}^1S_0$ channel.

\section{$H$ dibaryon (singlet $S$-wave)}
\label{sec:Hdib}

The $H$ dibaryon is a conjectured $uuddss$ bound state that is a
scalar and an $SU(3)$ singlet~\cite{Jaffe:1976yi}. Lattice
calculations performed with dynamical fermions agree that this bound
state exists for heavier-than-physical quark masses but do not agree
not on its binding energy, with results varying from a few MeV up to
75 MeV.

\begin{figure}
  \centering
  \includegraphics[width=\textwidth]{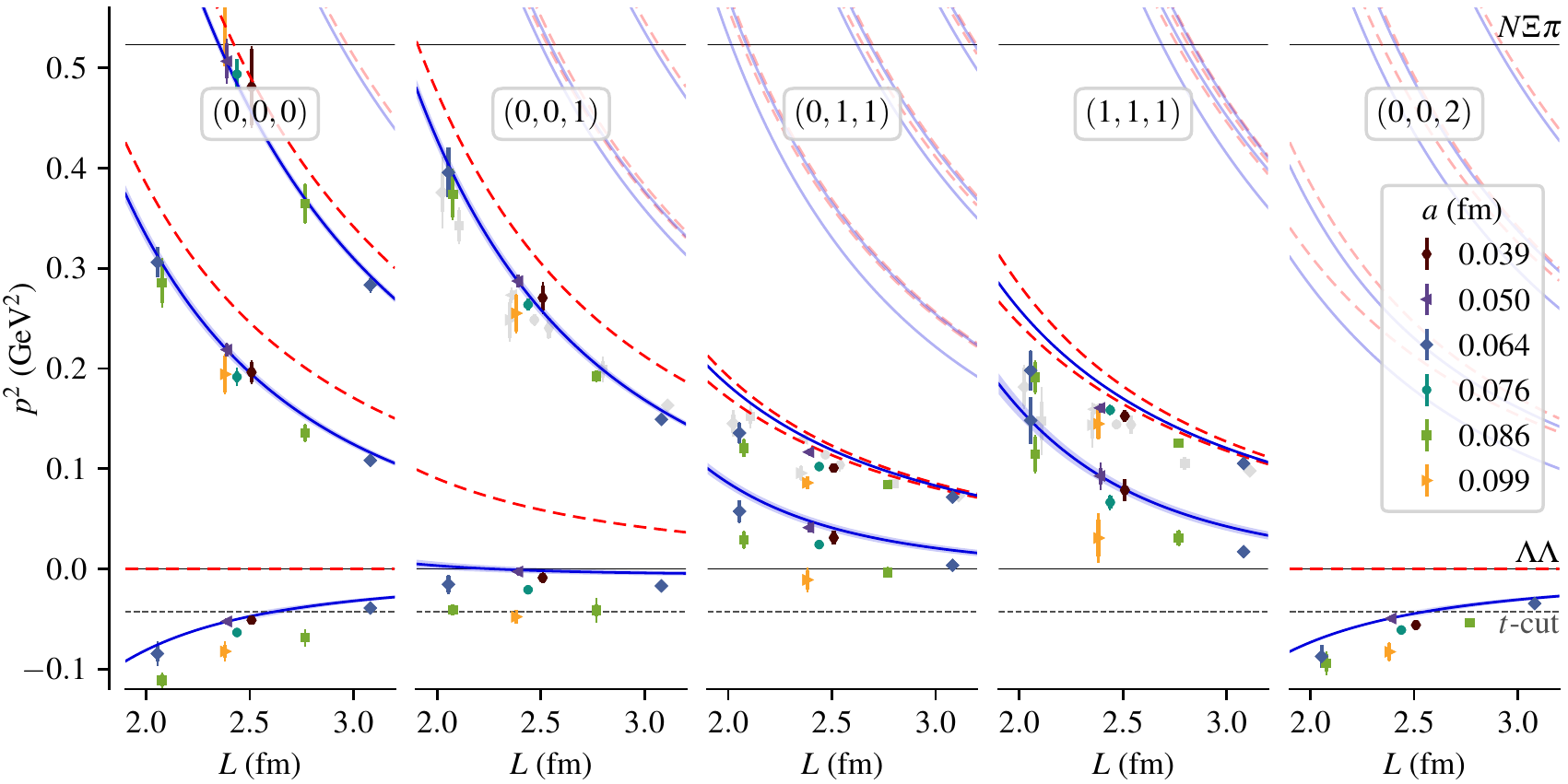}
  \caption{Singlet spectrum in trivial irreps: $p^2$ versus $L$ in the
    rest frame and four moving frames. The colored and gray points
    show lattice spin-zero and spin-one energy levels,
    respectively. The blue curves show the energy levels in the
    continuum obtained from a global fit and the red dashed curves
    show the noninteracting levels.}
  \label{fig:p2cm_H}
\end{figure}

The SU(3) singlet baryon-baryon spectra from our calculation, in the
trivial irrep of the rest frame and in four moving frames and on all
ensembles, are shown in Fig.~\ref{fig:p2cm_H}. One can see a clear
trend as the lattice spacing is varied, with coarser lattice spacings
corresponding to lower energies that are further below the
noninteracting levels.

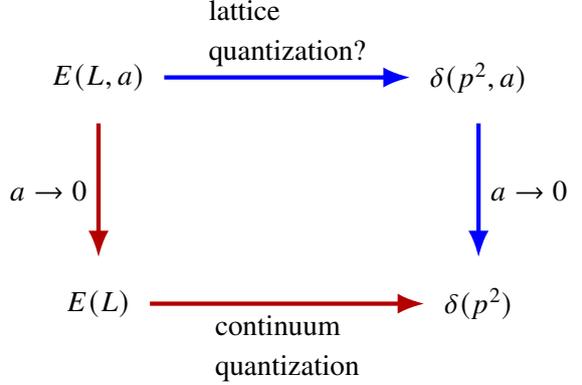
\begin{wrapfigure}{r}{0.5\textwidth}
  \begin{tikzpicture}[
    shorten < =  1mm, shorten > = 1mm,
node distance = 50mm, on grid, auto,
every path/.style = {-Latex,ultra thick},
sx+/.style = {xshift=1 mm},
sy+/.style = {yshift=1 mm},
sx-/.style = {xshift=-1 mm},
sy-/.style = {yshift=-1 mm},
p1/.append style = {draw=red},
state/.append style = {rectangle,draw=none},
                    ]
\node[state] (A) {$E(L,a)$};
\node[state] (B) [right=of A] {$\delta(p^2,a)$};
\node[state] (C) [below=30mm of A] {$E(L)$};
\node[state] (D) [right=of C] {$\delta(p^2)$};
\path[->,blue] %
 (A) edge node[above,black,align=left] {lattice\\quantization?} (B)
 (B) edge node[right,black] {$a\to 0$} (D);
\path[->,red!70!black] %
 (A) edge node[left,black] {$a\to 0$} (C)
 (C) edge node[below,black,align=left] {continuum\\quantization} (D);
\end{tikzpicture}
\caption{Two paths, red and blue, from the lattice finite-volume
  energy levels $E(L,a)$ to the continuum phase shift $\delta(p^2)$.}
\label{fig:paths}
\end{wrapfigure}

Given the energy levels at nonzero lattice spacing, our goal is to
obtain the continuum phase shift. Two strategies for this are
illustrated in Fig.~\ref{fig:paths}. Conceptually, it is most
straightforward to follow
the red path, by performing continuum extrapolations of the
finite-volume energy levels and then obtain the scattering amplitude
using quantization conditions, which have been derived in the
continuum. However, in practice this is difficult because it requires
matched volumes at multiple lattice spacings. Alternatively, one can
follow the blue path by first obtaining a scattering amplitude at
finite lattice spacing and then extrapolating it to the
continuum. However, this requires a quantization condition at finite
lattice spacing, which has only been studied for a simple model in
Ref.~\cite{Korber:2019cuq}.

Our strategy is to defer a rigorous understanding of quantization
conditions and scattering amplitudes at finite lattice spacing to
future work. Instead, we apply continuum quantization conditions to
data at nonzero lattice spacing and assume that symmetry-breaking
effects are small so that the effect of lattice artifacts is to only
modify the parameters of a continuum scattering amplitude.

Truncated to $S$-wave, the quantization condition becomes
\begin{equation}
  p\cot\delta_0(p^2) = B_{00}(p^2) \equiv \frac{2}{\sqrt{\pi} L \gamma}
  Z_{00}^{\vec P L/(2\pi)} \left(1, \left(\frac{pL}{2\pi}\right)^2 \right),
\end{equation}
where $Z_{00}^{\vec D}$ is a generalized zeta function. Given an
energy level corresponding to momentum $p^2$, this provides the phase
shift $\delta_0(p^2)$ at that scattering momentum. Conversely, given
an ansatz for $\delta_0(p^2)$, the solutions to this equation provide
the finite volume spectrum.

We determine $\delta_0(p^2)$ in the continuum by performing global
fits to the spectra from all of our ensembles. Our fit ansatz assumes
that $p\cot\delta_0(p^2)$ can be described by a polynomial in $p^2$,
\begin{equation}
  p\cot\delta_0(p^2) = \sum_{i=0}^{N-1} c_i p^{2i},\quad
  c_i = c_{i0} + c_{i1}a^2.
\end{equation}
We use the quantization condition to transform this ansatz for
$\delta_0(p^2)$ into an ansatz for the spectrum, and fit to the
spectrum. States below the $t$-channel cut or above the inelastic cut
are excluded from our fits. In addition, we exclude the excited states
in frames $(0,1,1)$ and $(1,1,1)$ because a nonzero $D$-wave amplitude
is needed to describe them (see the next section).

\begin{SCfigure}
  \centering
  \includegraphics{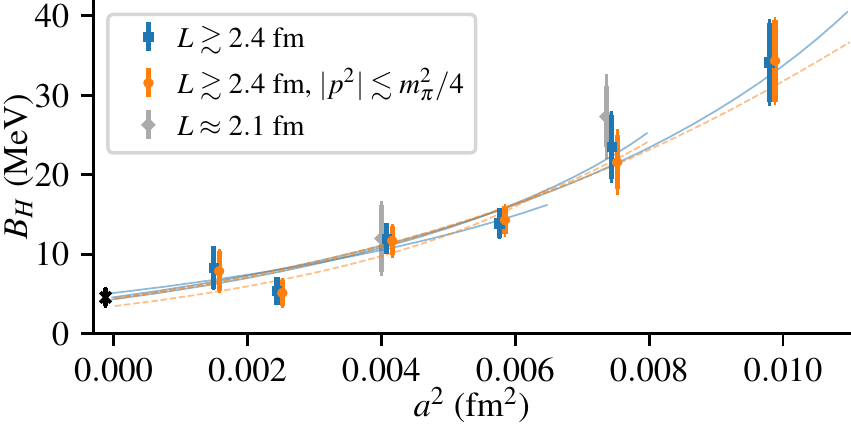}
  \caption{Binding energy of the $H$ dibaryon versus squared lattice
    spacing. The points are obtained from analyzing individual
    ensembles and the curves are from global fits to the spectra on
    multiple ensembles.}
  \label{fig:BH_vs_a2}
\end{SCfigure}

Bound states correspond to poles on the physical sheet of the
scattering amplitude below threshold, i.e.\ for $p=i\kappa$,
$\kappa>0$. They can be found by solving for
$p\cot\delta(p)=-\sqrt{-p^2}$. In all cases, we find a bound $H$
dibaryon. The binding energy depends strongly on the lattice spacing,
as shown in Fig.~\ref{fig:BH_vs_a2}. Our final result is
$B_H=4.56 \pm 1.13 \pm 0.63$~MeV, where the second (systematic)
uncertainty is obtained by varying the plateau region used to obtain
the spectra and by performing cuts on $a$, $L$, and $p^2$. For further
details about the analysis, including a cross-check based on the red
path in Fig.~\ref{fig:paths}, see Ref.~\cite{Green:2021qol}.

\section{Singlet $D$-wave (preliminary)}
\label{sec:Dwave}

\begin{figure}
  \centering
  \includegraphics[width=\textwidth]{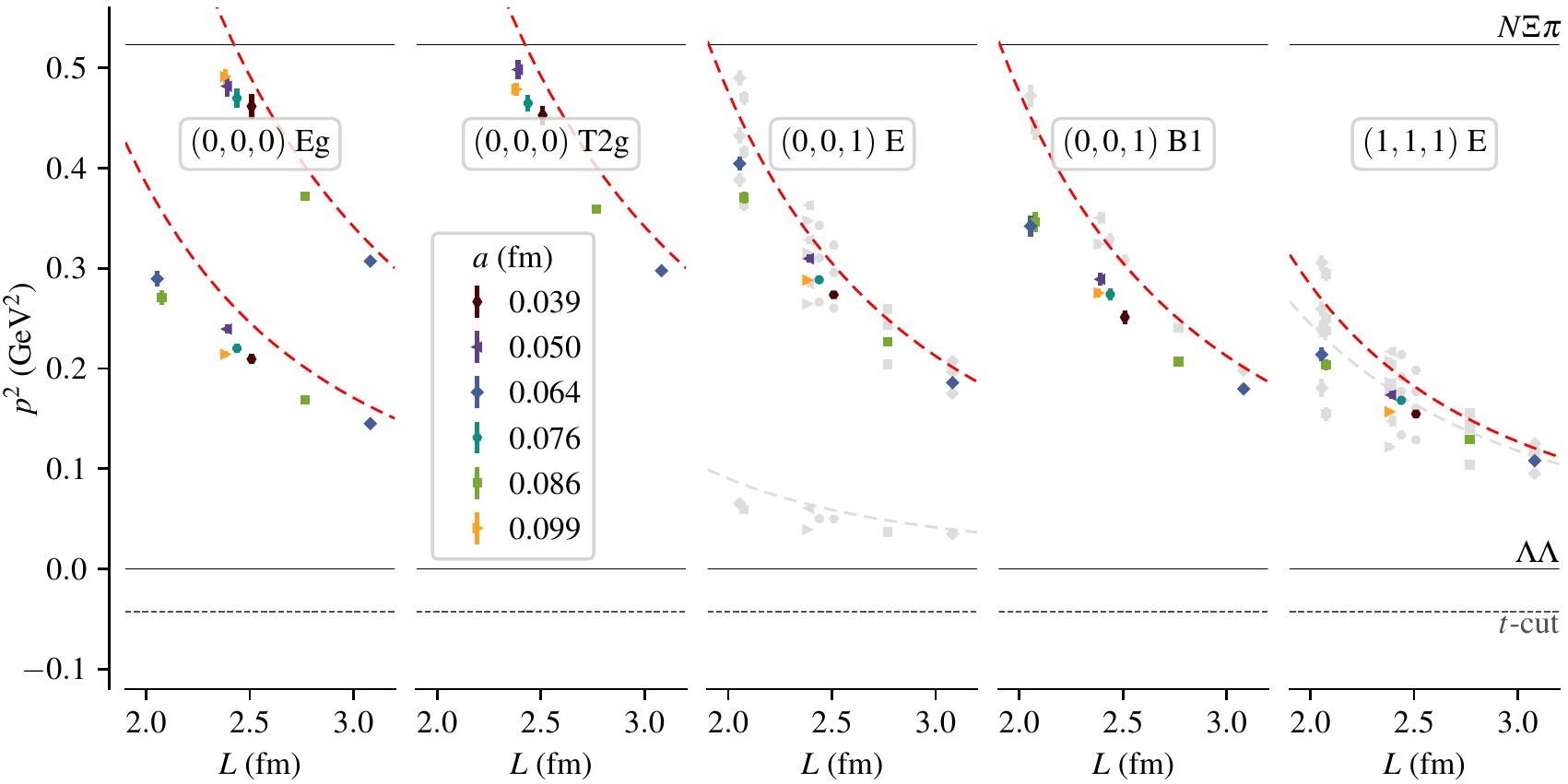}
  \caption{Singlet spectrum in nontrivial irreps containing spin-zero
    states. Spin-one states and spin-one noninteracting levels are
    shown in gray; see the caption of Fig.~\ref{fig:p2cm_H}.}
  \label{fig:p2cm_D}
\end{figure}

Nontrivial irreps provide information about higher partial
waves. Figure~\ref{fig:p2cm_D} shows preliminary estimates of SU(3)
singlet energy levels relevant for the ${}^1D_2$ partial wave. Here we
rely on overlaps between states and interpolating operators to
identify the spin-zero states; in some cases, there are many more
spin-one states.

For this flavor channel, both $\tilde K$ and $B$ are diagonal in spin,
so that the quantization condition factorizes and spin zero can be
analyzed independently of spin one. Neglecting $G$-wave and higher
partial waves, the quantization condition for the irreps in
Fig.~\ref{fig:p2cm_D} has the form
\begin{equation}
  \det\left[ p^5 \cot\delta_2(p^2)\, I_{n\times n} - B(p^2) \right] = 0,
\end{equation}
where $I$ is the $n\times n$ identity matrix, $B(p^2)$ is a matrix
involving generalized zeta functions, and $n=1$ or $2$. Given an
energy level corresponding to momentum $p^2$, the eigenvalues of
$B(p^2)$ provide the one or two possible values of
$p^5\cot\delta_2(p^2)$.

We also return to the trivial A1 irrep in frames $(0,1,1)$ and
$(1,1,1)$, for which we were unable to describe the excited-state
energy using only $S$ wave. Including both $S$ and $D$ waves, we get
\begin{equation}
  \det\left[
    \begin{pmatrix}
      p\cot\delta_0(p^2) & 0_{1\times n} \\
      0_{n\times 1} & p^5\cot\delta_2(p^2)\, I_{n\times n}
    \end{pmatrix} -
    \begin{pmatrix}
      B_{00}(p^2) & B_{02}(p^2) \\
      B_{20}(p^2) & B_{22}(p^2)
    \end{pmatrix}
  \right] = 0.
\end{equation}
Taking the Schur complement, we obtain
\begin{equation}\label{eq:d0_schur}
  p\cot\delta_0(p^2) = B_{00}(p^2) + B_{02}(p^2)
  \left[ p^5\cot\delta_2(p^2) I_{n\times n} - B_{22}(p^2) \right]^{-1} B_{20}(p^2).
\end{equation}
The fact that the low-lying spin-zero noninteracting levels are singly
degenerate implies that the matrix $B(p^2)$ has just one divergent
eigenvalue at each of these levels. From this, we find that the RHS of
Eq.~\eqref{eq:d0_schur} has a pole only when
$\det[p^5\cot\delta_2(p^2)I_{n\times n} - B_{22}(p^2)]=0$. If
$\delta_2(p^2)=0$, this reduces to the noninteracting levels. In
general, if $\delta_0(p^2)$ does not pass through zero, then an
interacting level will be found between every pair of poles. As seen
in Fig.~\ref{fig:p2cm_H}, describing our data requires that the poles
be shifted from the noninteracting levels, which implies a nonzero
$\delta_2(p^2)$.

\begin{SCfigure}
  \centering
  \includegraphics[width=0.6\textwidth]{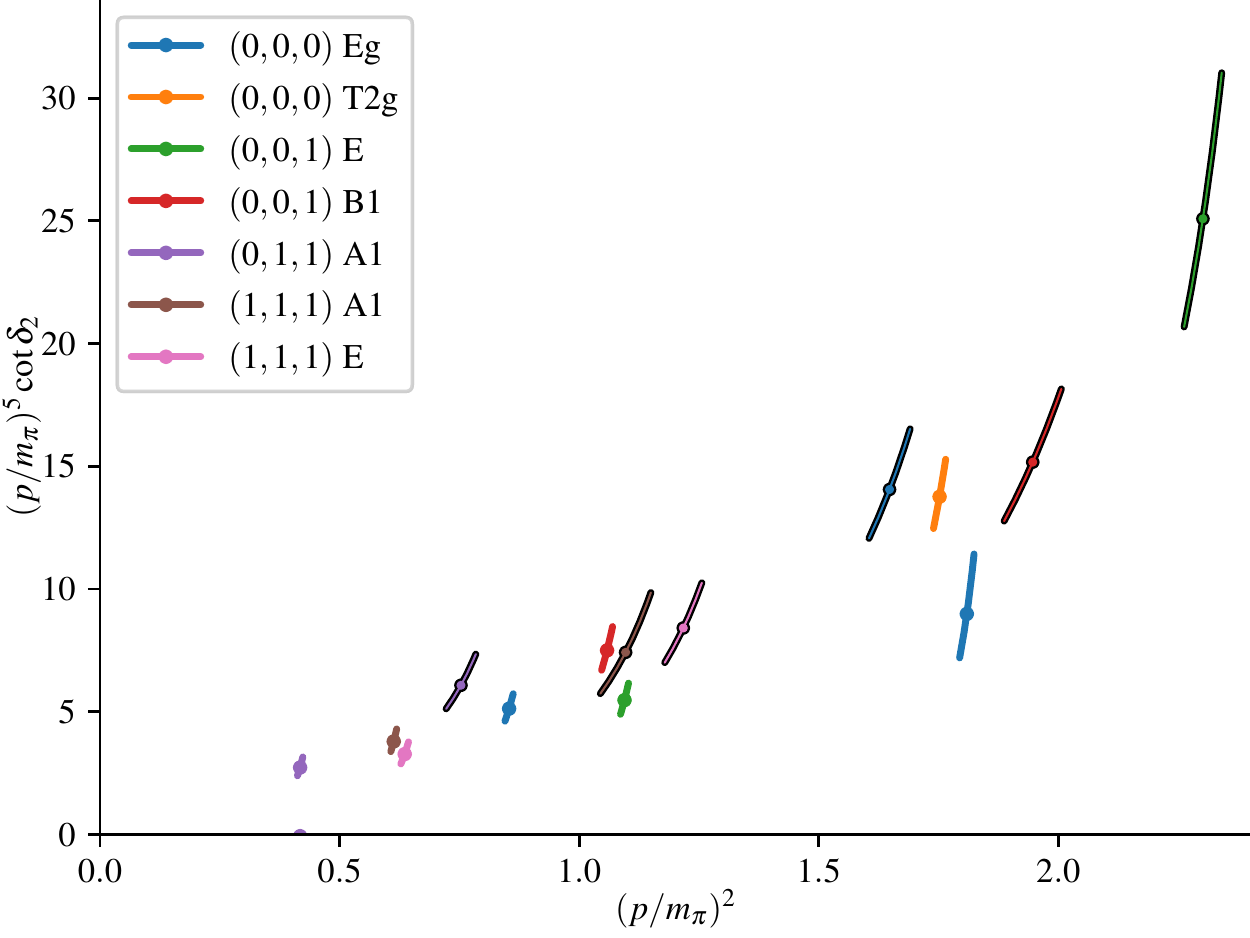}
  \caption{Singlet $p^5\cot\delta_2(p^2)$ versus $p^2$, in units of
    the pion mass, for ensembles N202 (solid colors) and H200 (with
    black outlines). For irreps with two solutions, just one of them
    appears within the bounds of this plot.}
  \label{fig:qc_D}
\end{SCfigure}

Taking the Schur complement in the opposite way, we find that the
possible values of $p^5\cot\delta_2(p^2)$ are given by eigenvalues of
\begin{equation}
  B_{22}(p^2) + \frac{B_{20}(p^2) B_{02}(p^2)}{p\cot\delta_0(p^2)-B_{00}(p^2)}.
\end{equation}
As a first study, we fix $\delta_0(p^2)$ based on fits to other levels
in trivial irreps. With this fixed, we find that the excited state in
frames $(0,1,1)$ and $(1,1,1)$ provides a good constraint on
$\delta_2(p^2)$. Preliminary results are shown in Fig.~\ref{fig:qc_D}
for two ensembles, N202 and H200, with the same lattice spacing but
different volumes. The $D$-wave phase shift obtained from these levels
is quite compatible with the phase shift obtained from nontrivial
irreps where $S$-wave does not contribute. Furthermore, these levels
are particularly useful because they have the smallest $p^2$ and
provide information closest to the threshold. We also note that a
nonrelativistic version of the analysis in
Ref.~\cite{Grabowska:2021xkp} could help to explain why these levels
are primarily influenced by $D$~wave.

\section{27-plet $S$-wave (preliminary)}
\label{sec:27}

The other SU(3) flavor irrep that appears only in the symmetric
product of two octet baryons is the 27-plet. This is particularly
interesting because it contains the nucleon-nucleon isospin-one
channel, where some calculations using a single lattice spacing have
reported a bound ``dineutron'' state~\cite{Yamazaki:2015asa,
  Berkowitz:2015eaa, Wagman:2017tmp, NPLQCD:2020lxg}. Here the
inelastic threshold is lower, corresponding to $m_B+m_D$, where $m_B$
and $m_D$ are the octet and decuplet baryon masses. Although the
octet-decuplet channel could be included in two-particle quantization
conditions, we do not include the relevant octet-decuplet
interpolating operators; therefore, our usable range of spectrum is
smaller in the 27-plet.

\begin{figure}
  \centering
  \includegraphics[width=\textwidth]{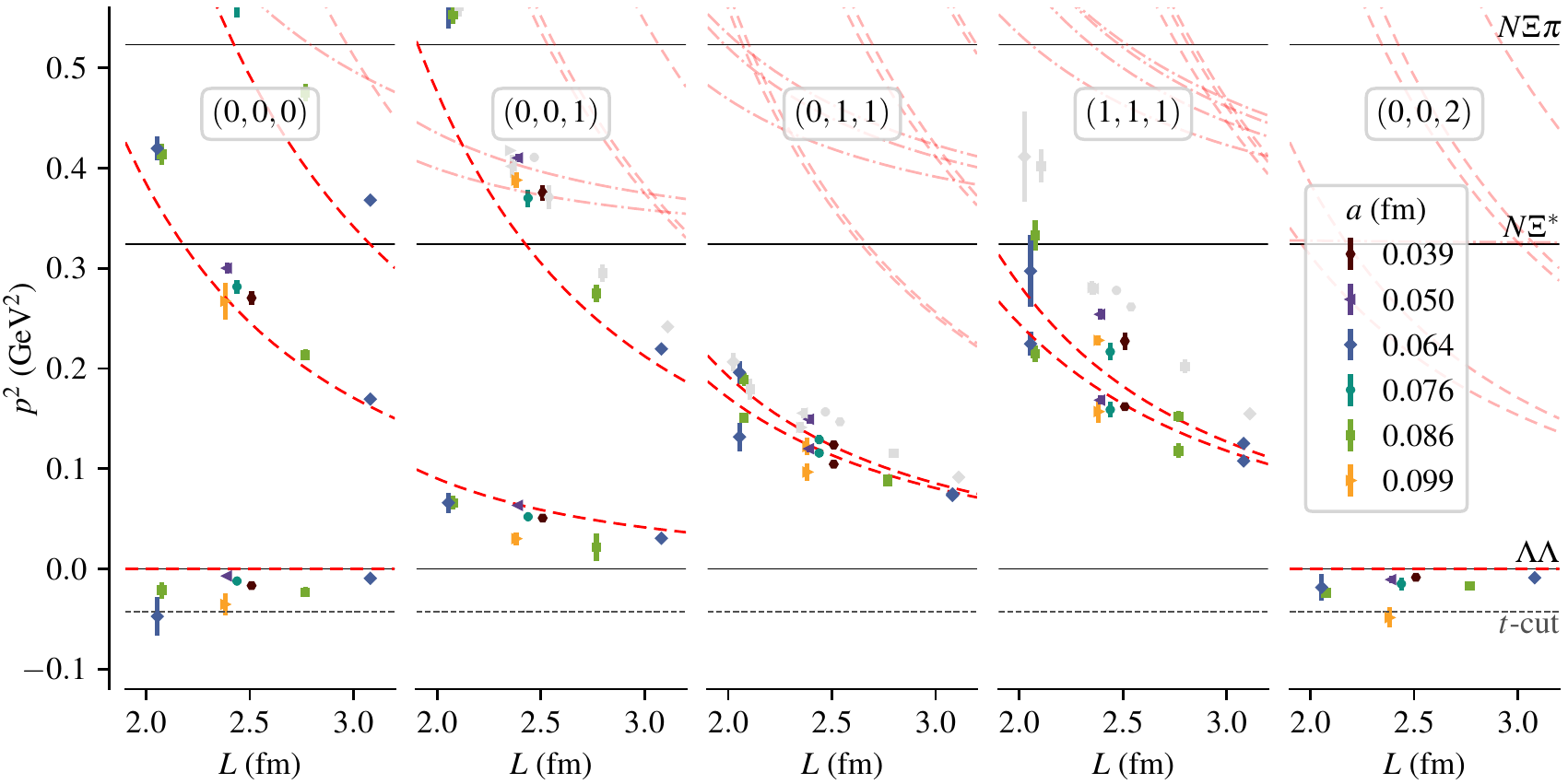}
  \caption{Septenvigintuplet spectrum in trivial irreps. In addition
    to the dashed curves showing noninteracting octet-octet levels, we
    also show dash-dotted curves depicting noninteracting
    octet-decuplet levels.}
  \label{fig:p2cm_27}
\end{figure}

Preliminary estimates of the spectra in trivial irreps are shown in
Fig.~\ref{fig:p2cm_27}. Near threshold, the downward shifts from
noninteracting levels are considerably smaller than in the singlet
sector, rendering a bound state rather unlikely. Higher in the
spectrum, the levels are shifted upward from the noninteracting ones;
this indicates that at some point the phase shift passes through
zero. Because of this, $p\cot\delta_0(p^2)$ cannot be well described
by polynomials, and we are exploring other fit forms such as rational
functions. The absolute size of the discretization effects is perhaps
a bit smaller than in the singlet sector; however, since the shifts
from noninteracting levels are also smaller, the relative effect on
the phase shift may still be large.

\section{Conclusions}
\label{sec:conclu}

Using distillation, we are able to determine the low-lying spectrum of
baryon-baryon states in a variety of frames, irreps, and flavor
channels, which can then be analyzed using finite-volume quantization
conditions. We have repeated this using several lattice ensembles with
six different lattice spacings, producing the first study of
discretization effects in a multibaryon system. Our results show a
strong dependence on the lattice spacing, which makes it essential to
include a continuum limit study: the binding energy of the
$H$~dibaryon on our coarsest lattice spacing is about 7.5 times larger
than in the continuum.

This work is currently being extended in two ways. First, we are
studying other systems at the SU(3)-symmetric point, in particular
nucleon-nucleon scattering. Second, we are studying the effect of
SU(3) breaking and the approach to the physical
point~\cite{Padmanath_proc}.

\small
\acknowledgments

We thank Dorota M.\ Grabowska, Maxwell T.\ Hansen, Ben Hörz, and Daniel Mohler for
helpful conversations.
Calculations for this project used resources on the supercomputers
JUQUEEN~\cite{juqueen}, JURECA~\cite{jureca}, and JUWELS~\cite{juwels}
at Jülich Supercomputing Centre (JSC). The authors gratefully
acknowledge the support of the John von Neumann Institute for
Computing and Gauss Centre for Supercomputing e.V.\
(\url{http://www.gauss-centre.eu}) for project HMZ21.
The raw distillation data were computed using
QDP++~\cite{Edwards:2004sx}, PRIMME~\cite{PRIMME}, and the deflated
SAP+GCR solver from openQCD~\cite{openQCD}. Contractions were
performed with a high-performance BLAS library using the Python
package opt\_einsum~\cite{opt_einsum}.  The correlator analysis was
done using SigMonD~\cite{sigmond}.  Much of the data handling and the
subsequent phase shift analysis was done using NumPy~\cite{numpy} and
SciPy~\cite{scipy}. The plots were prepared using
Matplotlib~\cite{Hunter:2007}. The quantization condition beyond
$S$-wave was computed using
TwoHadronsInBox~\cite{Morningstar:2017spu}.
This research was partly supported by Deutsche Forschungsgemeinschaft
(DFG, German Research Foundation) through the Cluster of Excellence
``Precision Physics, Fundamental Interactions and Structure of
Matter'' (PRISMA+ EXC 2118/1) funded by the DFG within the German
Excellence Strategy (Project ID 39083149), as well as the
Collaborative Research Centers SFB 1044 ``The low-energy frontier of
the Standard Model'' and CRC-TR 211 ``Strong-interaction matter under
extreme conditions'' (Project ID 315477589 -- TRR 211).  ADH is
supported by: (i) The U.S. Department of Energy, Office of Science,
Office of Nuclear Physics through the Contract No. DE-SC0012704
(S.M.); (ii) The U.S. Department of Energy, Office of Science, Office
of Nuclear Physics and Office of Advanced Scientific Computing
Research, within the framework of Scientific Discovery through Advance
Computing (SciDAC) award Computing the Properties of Matter with
Leadership Computing Resources. JRG acknowledges support from the
Simons Foundation through the Simons Bridge for Postdoctoral
Fellowships scheme. We are grateful to our colleagues within the CLS
initiative for sharing ensembles.

\bibliographystyle{JHEP}
\bibliography{refs}

\end{document}